# Illustrations of Equivalent Methods to Reproduce Vehicle and Occupant Dynamics as a Pedagogical Tool

Bob J. Scurlock, Ph.D., ACTAR and James R. Ipser, Ph.D.
*Department of Physics, University of Florida, Gainesville, Florida*

**Introduction**

When explaining to a lay audience the magnitude of forces or accelerations imparted to vehicles or experienced by vehicle occupants during a motor vehicle collision, it is often helpful to recast the critical results in terms of other physical systems or impact configurations which will reproduce the equivalent dynamics of the subject accident to serve as a conceptual aid for the audience. In this article, we present the basis for such equivalents and explicitly demonstrate, using two physics simulation software packages, that such equivalents are based on nothing more than the application of the laws of physics.

**Basic Collision Dynamics**

In a well-known result of collision dynamics, the maximum inelastic energy available during impact is directly proportional to the impact closing-speed rather than the velocities as measured with respect to any particular reference frame [1, 2, 3]. For a simple 1-D collision, the maximum inelastic energy is expressed by:

$$E^{Max} = \frac{1}{2} \cdot \bar{m}(v_{Rel,i})^2$$

where $\bar{m}$ is the system reduced mass and $v_{Rel,i}$ relative velocity, or closing-speed, at the moment of impact.

In addition to the relative-velocity at impact, the particular time-dependent forces exchanged by two colliding objects are dependent upon the specific geometrical properties of the impact, such as the location of contact and the moments of inertia, as well as the responses of the surfaces of contact to deflection and even deflection rate. As a consequence of the Galilean Invariance of Newton's laws, for free objects undergoing collision, the contact force can only depend upon the relative impact velocity (the closing-speed) rather than the velocities measured with respect to any one reference frame (such as the Earth-fixed frame) [4]. This implies, when one considers an ensemble of possible impact scenarios between two vehicles, neglecting tire forces, each vehicle is guaranteed to experience the same unique acceleration independent of the ground speeds of the vehicles so long as the impact configurations remain the same and the impact closing-speeds are the same. This relation is true for any two bodies undergoing impact, whether they are two colliding vehicles or the subsequent impact of an occupant whose body collides with his seat.

**A Simple EDSMAC4 Simulation**

In order to support the above stated consequence of Newtonian physics, we staged three front-to-rear impacts using the EDSMAC4 [5] simulator in the HVE software package from Engineering Dynamics Corporation [6]. Let us consider the subject accident a rear-end collision between a Lincoln Navigator and an Acura TL. Suppose the Lincoln Navigator impacts the rear of the Acura with a speed of 5 mph, while the Acura was initially at rest. During this impact of course, the Acura will undergo a forward acceleration, while the Navigator will undergo a deceleration. Suppose in an effort to explain to a lay audience the magnitude of the acceleration imparted to the Acura, we wanted to recast the acceleration estimated in this first scenario (the subject accident) in terms of equivalent ways of reproducing the impact forces on the Acura. From the prior section, we know we can reproduce the identical impact forces so long as the impact configuration and closing-speed remain the same. Since the closing-speed is simply the difference of the impact velocities of the two vehicles, there are an infinite number of ways to collide the two vehicles such that the closing-speed remains exactly equal to 5 mph. To illustrate our point, we focus on two other impact scenarios. In our second simulated scenario, we allow the Lincoln Navigator to move forward at 2.5 mph while the Acura moves backward at 2.5 mph. This again results in a closing-speed equal to 5 mph. In our third scenario, we hold the Lincoln at rest while letting the Acura move backward at 5 mph into the Lincoln. Figure 1 shows the velocity versus time for all three scenarios. As expected, the velocity change is identical in all three cases, as momentum conservation tells us that the change-in-velocity is dependent upon the relative impact speed (closing-speed). We also note the slopes of the three lines are nearly identical. This is an indication that the impact forces are equivalent in the three scenarios.

Figure 2 shows the resulting acceleration pulses as a function of time on the Acura for all three scenarios. We can see from this graph the acceleration pulses are nearly identical in all three cases as expected from Newton's laws. If we wanted to explain to a lay person how one can reproduce the vehicle impact forces involved in the subject accident, we can confidently point to scenario 3 for example, whereby we have our audience imagine the Acura rolling backward into the Lincoln at 5 mph. As demonstrated by the simulation, this is guaranteed to give the same vehicle impact forces. Anyone who has never been rear-ended, but has had the misfortune of backing into another vehicle, may then be able to fully appreciate the level of forces involved in the subject accident.

**Consequences of the Work-Energy Theorem**

Another useful way to convey the magnitude of impact speeds to a lay audience is to recast these estimates into equivalent ways of reproducing them by using unrelated mechanical systems such as swings (pendulums) and slides (inclined planes). This is particularly true of low-speed collisions, as one generally has a better feel for higher vehicle speeds than lower vehicle speeds. For example, to say a vehicle impact occurred with a closing-speed of 3 mph may give a different impression to a lay person compared to saying it occurred at the average human walking speed. For demonstrative purposes, setting such magnitudes into new and creative contexts by using equivalent mechanical systems can be quite a powerful pedagogical tool.

The work-energy theorem tells us a conservative force field (such as gravity) will accelerate an object (such as a vehicle) in a way that is independent of the particular path taken (sled versus pendulum) and only dependent upon the distance traveled along the direction of the field itself [7]. Formally, this is given by the expression:

$$\int_1^2 \bar{F} \cdot d\bar{r} = U_1 - U_2$$
$$= mgH_1 - mgH_2$$
$$= \frac{1}{2}m(v_2^2 - v_1^2)$$

or rewritten in its usual form, we have:

$$2g(H_1 - H_2) = v_2^2 - v_1^2$$

where $g$ is the acceleration due to Earth's gravity, $m$ is the mass of the object, $U_1$ represents initial gravitational potential energy ($mgH_1$), and $U_2$ represent the final gravitational potential energy ($mgH_2$) for an object being displaced from point 1 to point 2. This change in potential energy must be equal in magnitude to the change in kinetic energy. This implies that an object sliding down a frictionless plane will have the same final velocity at the end of its path as an object swinging from a rope that reaches the bottom of its arc, so long as in the two cases both objects undergo the same change in height from start to finish. This means, if we estimate the speed of object *A* prior to impacting object *B* (say two cars), we can illustrate a way of reproducing the impact speed of object *A* by imagining object *A* first sliding down a frictionless inclined plane by the appropriate height to yield the correct pre-impact speed. We can also imagine object *A* swinging as a pendulum until it reaches the bottom of its arc, where impact occurs just at that moment (see Appendix). Again assuming the total vertical distance traveled is the same as that in the inclined plane example, the final pre-impact speed is guaranteed to be identical by the work-energy theorem.

As a consequence of conservative nature of Earth's gravitational field, the automobile industry has a long history, going back many decades, of using sled and pendulum test devices within various laboratories around the world. Indeed, even a casual search in the Society of Automotive Engineers journal reveals literally thousands of articles in which the words "sled" and "pendulum" appear [8]. One can also find many biomechanics studies conducted where volunteers or cadavers have been gravitationally accelerated down inclined planes in order to reach the needed impact closing-speeds for various tests [see for example 9, 10, 11].

Pendulum impact devices are used to test the strength of vehicle components such as bumper systems, crash dummy response, and even the strength of human tissues [see for example 12, 13, 14]. The reason these sled and pendulum devices are routinely used is given by the work-energy theorem itself. Independent of the particular path taken, Earth's gravitational field will always accelerate these objects such that their velocities are only dependent upon the total vertical distance traveled. This means, in a laboratory setting, one can precisely control the required impact speeds simply by adjusting the starting height of the test apparatus. Also, because sleds and pendulums do not require complex rigs or motors to accelerate the test devices, a test apparatus can be easily contained within the confines of a small laboratory space.

**An Explicit Demonstration of the Work-Energy Theorem using Physics Simulators**

The consequences of the work-energy theorem can be derived in a few lines of algebra. As a way of explicitly demonstrating the result, we conducted a series of sled simulations using the SIMON simulator in HVE [15] as well as the Articulated Total Body (ATB) simulator [16]. As a first test of the work-energy theorem, we simulated an Acura TL sliding backward down a frictionless inclined plane with a 5 degree slope. The Acura's suspension and tires were effectively made rigid within HVE in order to approximate ideal sled test conditions. Figure 3 shows the starting configuration of the sled test. Here we let the Acura slide down the sled such that the vertical distance traveled was 13.75 ft.

Figure 4 shows the resulting velocity of the Acura as a function of vertical distance traveled down the inclined plane. Using the Acura's center-of-gravity position as a function of time output by HVE, we were able to estimate the expected velocity by simply applying the work-energy theorem, where the velocity equals $\sqrt{2g\Delta H}$. Here, $\Delta H$ is the magnitude of the change-in-height of the vehicle's center-of-gravity. The resulting ideal sled estimate, based on work-energy, yields a nearly identical result to the HVE physics simulation. Indeed we note the differences are at the level of 1% or less.

The Articulated Total Body package is a general purpose 3-D physics simulator that can be used to simulate a wide range of physical systems, including vehicle collisions and occupant dynamics. Within ATB, we constructed a simple sled device to slide down a 10 degree frictionless plane such that the total vertical height changes by the same magnitude (13.75 ft) as in our HVE simulation. Figure 5 shows the ATB sled and frictionless plane. The resulting sled velocity versus the magnitude of the vertical change-in-height is shown in Figure 6. The center-of-gravity positions of the sled object as reported by ATB were used to estimate the expected velocity again by using the work-energy theorem. As with the HVE simulation, we see excellent agreement between the ATB simulator and the work-energy theorem to better than 0.5%. This again demonstrates the validity of the work-energy theorem as is expected, since it is simply a consequence of Newton's laws.

As a final demonstration of the work-energy theorem, we sampled the total final velocities (the velocity after reaching the flat surface) as a function of total centers-of-gravity changes-in-height for nine starting points on the inclined plane within the HVE simulator. The result is shown in Figure 7. We again compare the result from the HVE simulation to our expectation by applying the work-energy theorem. The results are in agreement to within 1%.

**Equivalent ways to Reproduce Occupant Forces using Equivalent Barrier Speed Impacts**

So far we have demonstrated the pedagogically useful ideas that (1) impact forces are dependent upon the impact closing-speeds and (2) vehicle-to-vehicle impact closing-speeds can be reproduced in laboratory conditions by taking advantage of the work-energy theorem and allowing one vehicle to gravitationally accelerate in a controlled way using inclined planes or pendulums such that the desired closing-speed is achieved.

Finally, we would like to demonstrate the equivalence between delivered occupant forces for two dissimilar collision systems. The first system (scenario 1) can again typically be thought of as the subject accident in which a Lincoln Aviator SUV impacts an Acura TL. Our second system (scenario 2) consists of an Acura TL sliding backward down a frictionless inclined plane such that it is allowed to impact an immovable rigid barrier. In this second scenario, the Acura achieves the barrier equivalent speed, or $V_{EBS}$, necessary to yield the identical delta-V as in the subject accident. The total changes-in-height needed to yield the appropriate $V_{EBS}$ speeds are depicted in Figure 7, and again show excellent agreement with our expectations from application of the work-energy theorem.

We used the HVE package with the DyMESH 3-D collision simulator [17] to estimate the acceleration pulses delivered to the Acura TL for both scenarios. As expected from basic collision physics, the two acceleration pulses exhibit differences in shape as these pulses are directly dependent upon the given system's reduced mass and effective stiffness; however, as shown below, both scenarios yield effectively the same occupant forces. We simulated these two scenarios such that a total change-in-velocity of the Acura of magnitudes 4, 5, and 6 mph was achieved in each case. Figure 8 shows the Acura $V_{EBS}$ values versus delta-V using the DyMESH simulator. An iterative process was used to adjust the Acura's $V_{EBS}$ value by varying the starting position along the frictionless plane, such that the desired delta-Vs were achieved. A fit to the data points reveals that the effective average restitution used by the DyMESH collision model in this closing-speed regime is of the order 22%.

In order to maximize the impact forces delivered to the Acura occupant, one can assume the Acura was fully accelerated such that it achieved its full delta-V prior to the occupant's body contacting the backrest. Taking the resulting acceleration pulses from HVE in each of the three test cases, we simulated the occupant motion assuming a 205 lb driver sitting in a frictionless chair (also maximizing contact forces) with an effective stiffness of 100 lbs/in. We separated the occupant from the backrest prior to impact such that first contact is made by the upper torso after the vehicle has already reached its full delta-V. This guarantees that the occupant contact forces, and therefore the impact severity to the occupant, will be maximized. The resulting upper torso velocities measured with respect to the vehicle are shown in Figures 9, 10, and 11 for the 4, 5, and 6 mph Acura delta-V tests respectively. We observe similar features in each test. That is, in each test, the vehicle-to-vehicle impact scenario yields the same Acura delta-V as the $V_{EBS}$ vehicle-to-barrier scenario. We note the different overall shapes of the vehicle velocity curves in both scenarios as is expected with differing acceleration pulse shapes; however, the torso velocity curves show very similar shapes. In particular, as the torso is accelerated from its maximum negative velocity (with respect to the vehicle) to common velocity (0 mph with respect to the vehicle), the overall shapes of the two curves are nearly identical, indicating that the forces on the occupant in either scenario are effectively the same to a good approximation. This is merely the same physics principle at work governing the behavior of colliding objects. The dynamics of colliding objects, holding all else the same, will only depend on the impact closing-speed.

Figure 12 shows the resulting average acceleration to the upper torso of the ATB-simulated occupant as a function of the vehicle delta-V for the two different scenarios over the ensemble of three tests [18]. The graph demonstrates that the average acceleration differs by less than a few percent for the two scenarios, independent of the Acura's total delta-V. Indeed, any discrepancies between the two scenarios likely have only to do with the exact starting time of the acceleration pulses simulated in ATB, as a minor difference in this starting time can allow the ATB-simulated occupant to relax into a slightly different posture at the moment the upper torso makes first contact with the backrest. Nevertheless, the principle of equivalents using the work-energy theorem and $V_{EBS}$ is clearly demonstrated.

**Conclusion**

We have demonstrated how one may make use of hypothetical sled and pendulum experimental devices to illustrate methods by which one can reproduce desired vehicle speeds through the simple application of the work-energy theorem. We have also demonstrated that the impact forces are determined by the relative impact speeds of two colliding bodies, and that, keeping all other conditions equal, an ensemble of collision scenarios will yield identical impact forces so long as the relative impact speeds are kept constant. Using both the equivalent scenarios to reproduce particular vehicle velocities guaranteed by the work-energy theorem and the equivalent collision scenarios, which yield identical occupant-cabin impact closing speeds, we have demonstrated that one can effectively reproduce forces experienced by the vehicle occupant in a rear-end impact by carefully selecting the appropriate closing-speeds for a barrier impact.

*Bob Scurlock, Ph. D. is a Research Associate at the University of Florida, Department of Physics and works as a consultant for the accident reconstruction and legal community. He can be reached at BobScurlockPhD@gmail.com. His website offering free analysis software can be found at: ScurlockPhD.com.*

*James Ipser, Ph. D. is Professor Emeritus at the University of Florida, Department of Physics. He regularly consults and provides expert opinion in the areas of vehicular accident reconstruction and biomechanical physics. He can be reached at JIpser@gmail.com. His website can be found at JIpsier.com.*

**References**

[1] "Rigorous Derivations of the Planar Impact Dynamics Equations in the Center-of-Mass Frame", Cornell University Library arXiv:1404.0250, B. Scurlock and J. Ipser.

[2] "Injury Causation Analyses: Case Studies and Data Sources", A. Damask and J. Damask, The Michie Company, Charlottesville, Virginia, 1990.

[3] "Accident Reconstruction", J. Collins, Charles C. Thomas, Springfield, Illinois.

[4] "Mechanics, Volume 1 of Course of Theoretical Physics", L. Landau and E. Lifshitz,

[5] "An Overview of the EDSMAC4 Collision Simulation Model", SAE 1999-01-0102, T. Day.

[6] Edccorp.com

[7] "Classical Dynamics of Particles and Systems", J. Marion and S. Thornton, Harcourt College Publishers, New York, New York, 1995.

[8] Search for "sled": http://www.sae.org/search/?qt=sled+&x=0&y=0. Search for "pendulum": http://www.sae.org/search/?qt=pendulum&x=0&y=0

[9] "Effect of Seat Stiffness in Out-of-Position Occupant Response in Rear-End Collisions", SAE 962434, B. Benson et al.

[10] "Effect of Seat Belts Equipped with Pretensioners on Rear Seat Adult Occupants in High-Severity Rear Impact", SAE 2008-01-1488, M. Tavakoli et al.

[11] "An Analysis of Traumatic Rupture of the Aorta in Side Impact Sled Tests", SAE 2005-01-0304, J. Cavanaugh et al.

[12] "Development of an Impact Pendulum for Use in Collinear, Low-Velocity Front-to-Rear Crash Tests", SAE 2006-01-1401, H. Guzman et al.

[13] "Airbag Bumpers Inflated Just Before the Crash", SAE 941051, C. Clark and W. Young.

[14] "Cadaver and Dummy Knee Impact Response", SAE 760799, J. Horsch and L. Patrick.

[15] "SIMON: A New Vehicle Simulation Model for Vehicle Design and Safety Research", SAE 2001-01-0503, T. Day, S. Roberts, and A. York.

[16] "The Effect of Inflated Backrest Stiffness on Shearing Loads Estimated with Articulated Total Body", Cornell University Library arXiv:1312.4097, B. Scurlock, J. Ipser, and P. Borsa.

[17] "The DyMESH Method for Three-Dimensional Multi-Vehicle Collision Simulation", SAE 1999-01-0104, T. Day and A. York.

[18] A video depicting the 5 mph delta-V simulations is available on the web: http://youtu.be/d1DbTTRLBY0

[19] "Use of ROOT in Vehicular Accident Reconstruction," Accident Reconstruction Journal, Vol. 21, No. 3, 2011, B. Scurlock.

[20] A video depicting this pendulum scenario is available on the web at: http://youtu.be/e8FqnMEqle4

[21] ARAS360.com

**Appendix**

**The Pendulum Equivalent**

Pendulum impact devices have been created and used by various laboratories for automobile and biomechanical testing. Though such devices can be difficult to simulate in packages such as ATB, the authors have created a simple C++ based pendulum model where Newton's laws are numerically integrated over time, and the corresponding pendulum motion path can be output. We know from the work-energy theorem that the final velocity of an object swinging or sliding down a trajectory is independent of the particular path followed by the object undergoing gravitational acceleration, but rather depends on the vertical distance traversed. This implies that so long as an object, such as a vehicle, is made to swing through the correct vertical distance, it can be made to have the correct final velocity so as to result in the required pre-impact speed for the given experiment.

Figure 13 shows a basic proof of principle of the C++ based pendulum simulator. This simulator is run in the ROOT data analysis package [19]. In the graph, we see the resulting angular position of a pendulum for massless rod length of 1 ft. The pendulum is pulled back and raised to a height of 0.4 in and released. It reaches a maximum velocity of 1 mph at the bottom of its arc. From basic classical mechanics, using the small angle approximation, we expect the angular frequency $\omega = \sqrt{g/l}$, where the time-dependent angular displacement is well approximated by the solution to the second-order differential equation [7]:

$$\ddot{\theta} + \omega^2 \theta = 0$$

Fitting to sinusoidal form of the pendulum's angular displacement, we can obtain an estimate for Earth's gravitational constant $g$ using the output data from the ROOT-based model. The fit parameter corresponding to ω was less than 1% different from $g$ at sea level in this case.

Figure 14 shows the result of pendulum simulation, where the mass traverses a 13 ft vertical distance [20]. Shown in red is an estimate of the velocity based on the work-energy theorem, where the mass's vertical position is sampled as a function of time. Here we have better than 0.1% agreement between the C++ simulation and the expected final velocity estimated by work-energy. The simulated pendulum motion can easily be sampled from our C++ based script and imported into a 3-D visualization tool such as ARAS 360 [21]. The rendered output is shown in Figure 15. These figures are captured from an animation sequence depicting a hypothetical pendulum apparatus, where the vehicle is made to swing through a vertical distance of 13 ft. We imagine the vehicle is allowed to impact a rigid barrier at the bottom of its arc. Imagining an occupant seated within this vehicle, we expect this scenario to faithfully reproduce the equivalent occupant forces as would be experienced in a rear-end impact, where the vehicle's delta-V is the same in both scenarios. Both the validity of using the work-energy theorem to obtain the appropriate equivalent swing height and the validity of equivalent occupant forces being derived by barrier impact are assumed in this illustration.

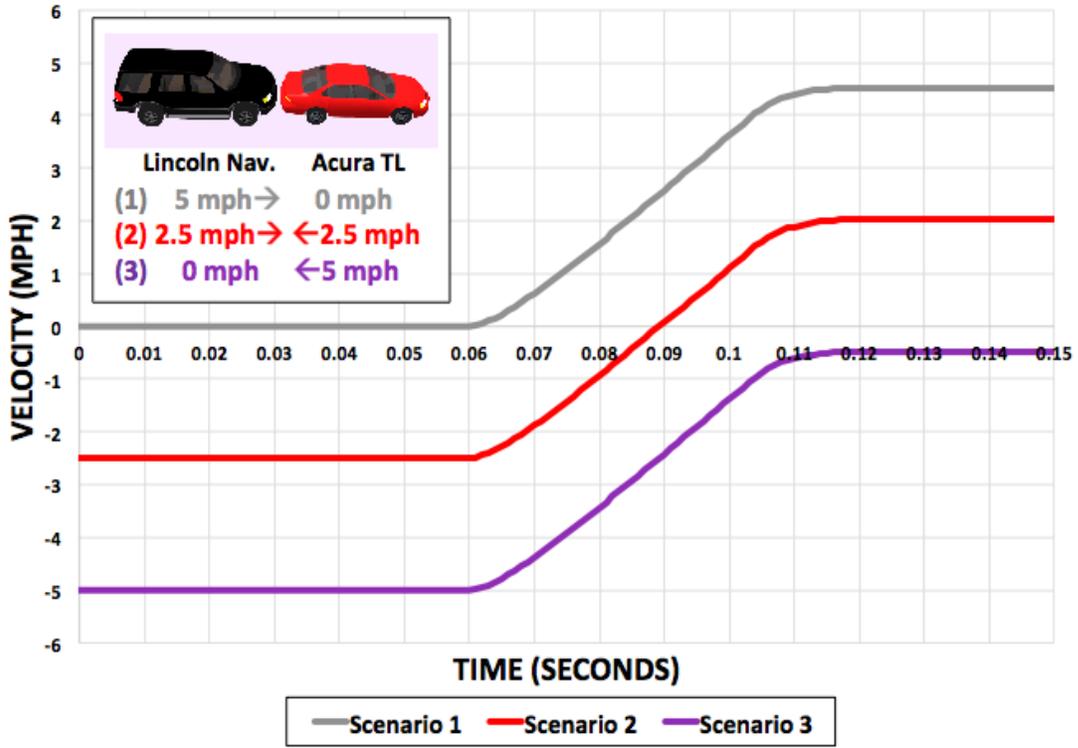

Figure 1: Velocity versus time for the three impact scenarios depicted in graphic.

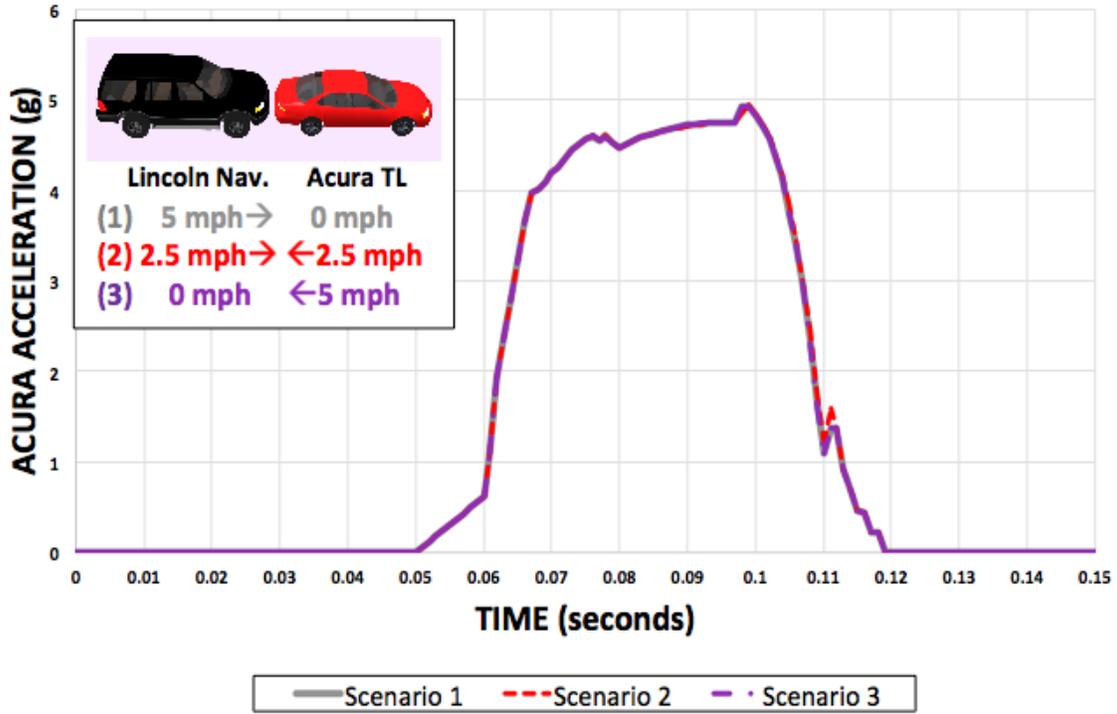

Figure 2: Acceleration versus time for the three scenarios depicted in graphic.

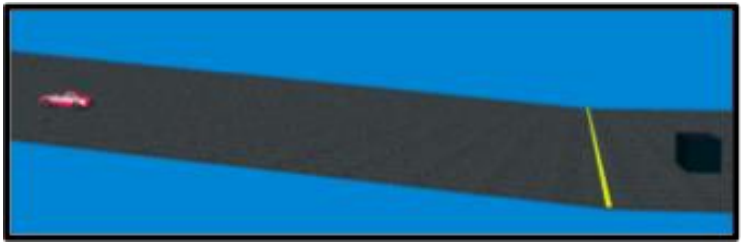

**Figure 3: Starting conditions for sled test. Acura is placed 13.75 ft above the flat surface and allowed to slide backward along a frictionless inclined plane.**

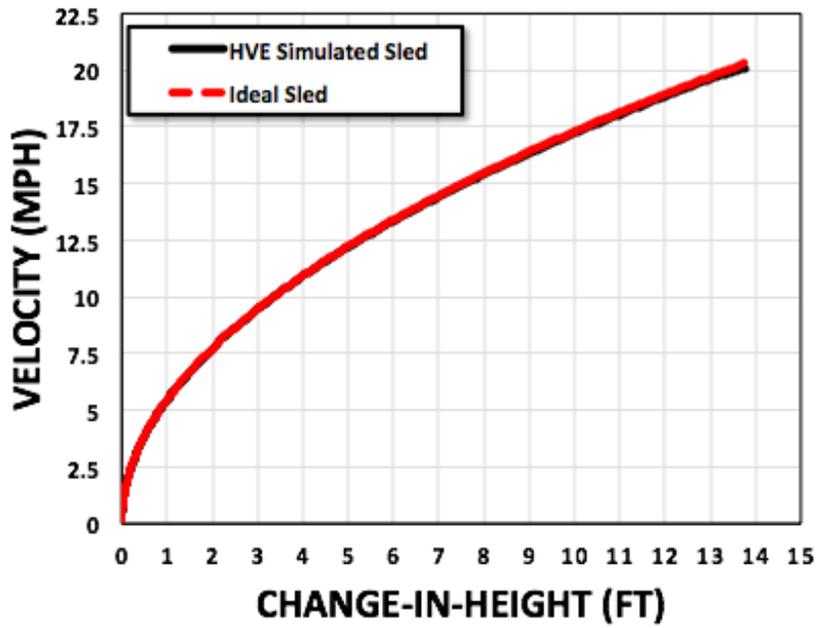

**Figure 4: Sled velocity as a function of change-in-height. The black curve shows the HVE-simulated velocity. Red shows the estimate based on work-energy.**

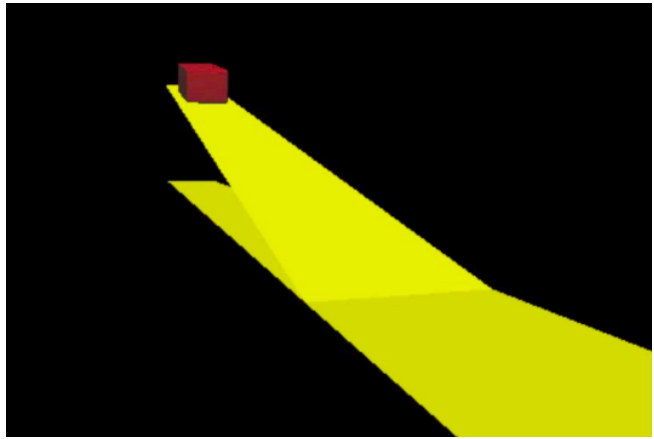

**Figure 5: Starting configuration of an ATB-simulated sled.**

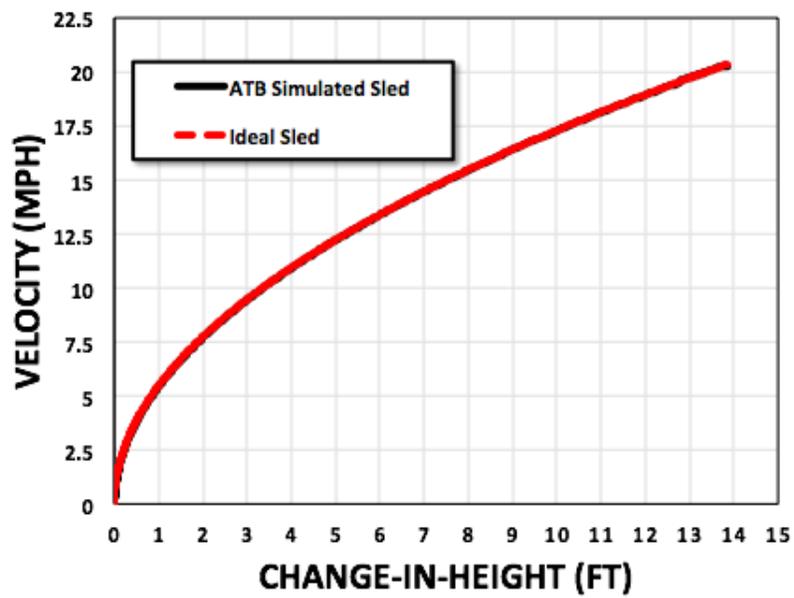

**Figure 6: Velocity versus change-in-height for an ATB-simulated sled travels a total vertical distance of 13.75 ft.**

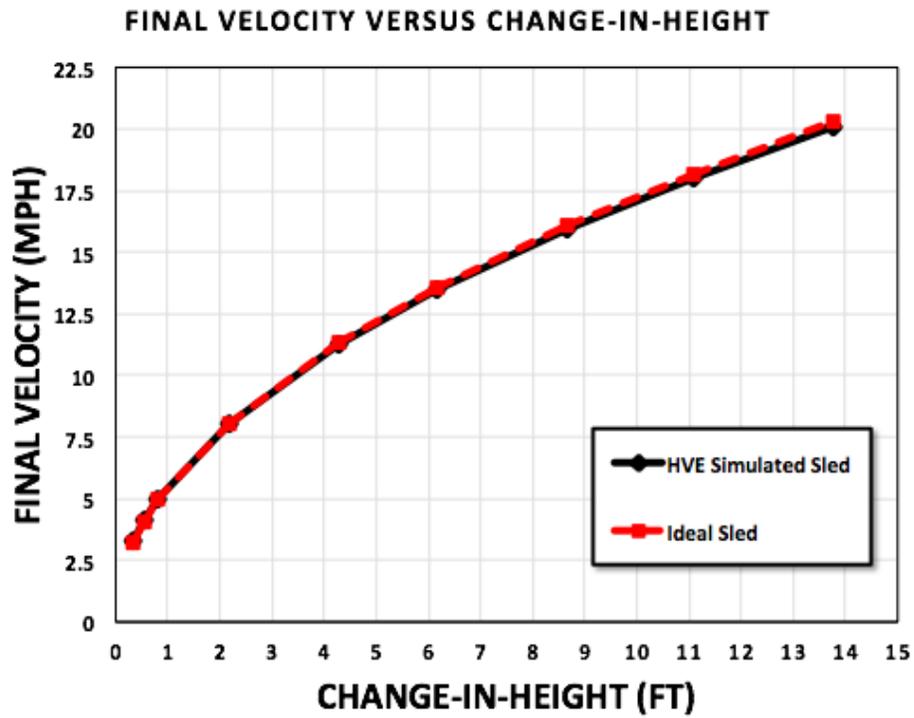

Figure 7: Final velocities versus total changes-in-height for an ensemble of nine sled tests conducted with the HVE simulator.

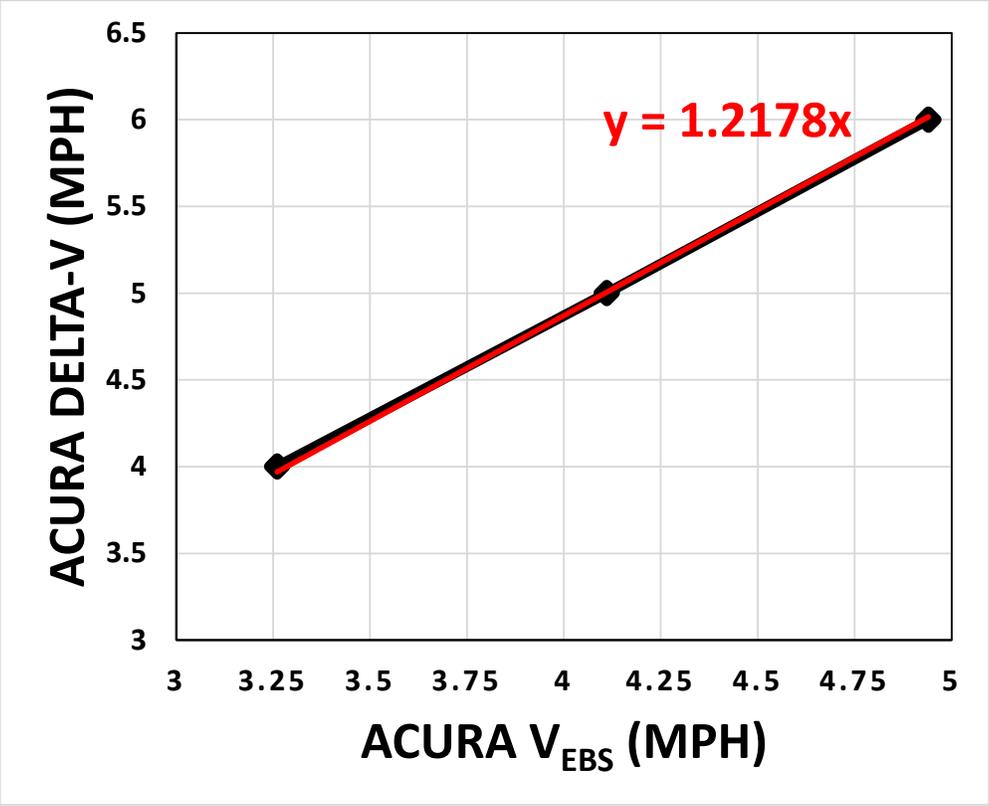

Figure 8: Black: Equivalent barrier speeds for Acura needed to reproduce desired delta-Vs. Red: The slope from a simple linear regression to the data points gives an estimate of the average restitution used by the DyMESH collision model, where we have $\varepsilon \approx 22\%$.

# 4 mph Delta-V Equivalents

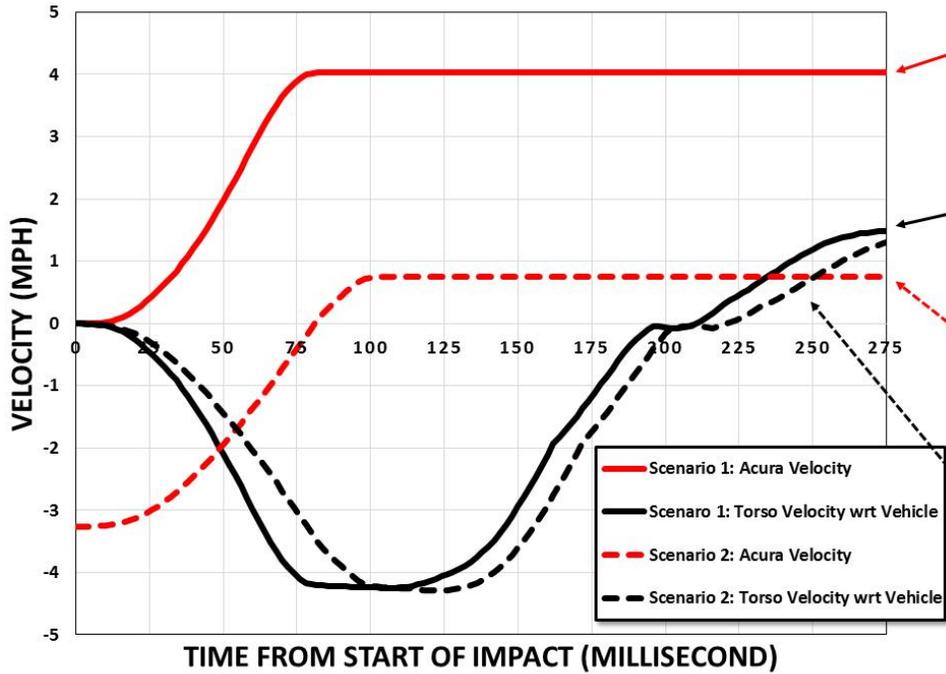
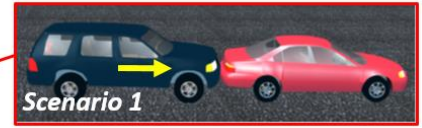
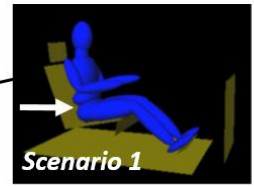
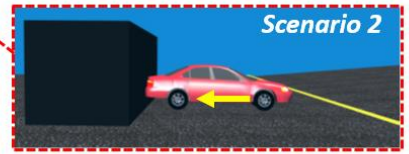
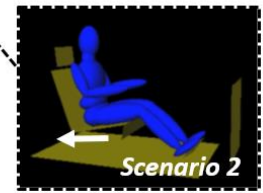

Figure 9: Vehicle and Occupant-Torso Velocities versus time from the start of impact. The vehicle-to-vehicle and vehicle-to-barrier simulations were generated such that the Acura experienced a change-in-velocity of 4 mph.

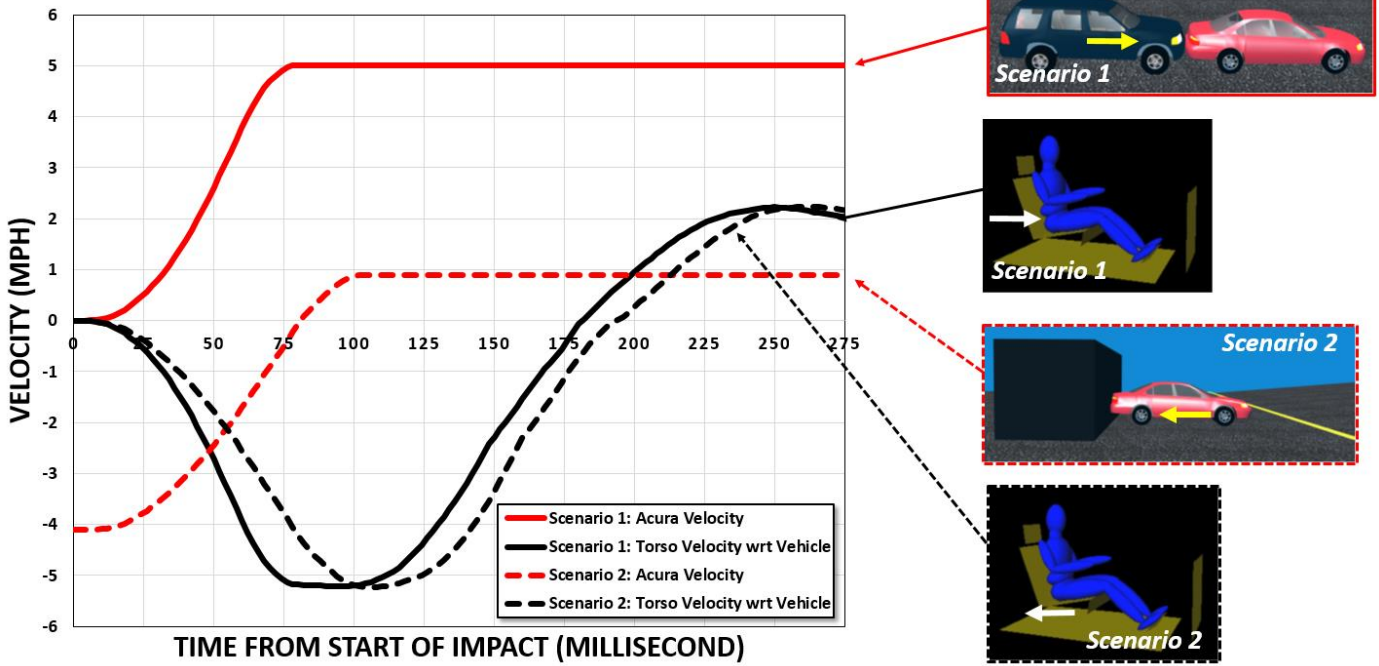

Figure 10: Vehicle and Occupant-Torso Velocities versus time from the start of impact. The vehicle-to-vehicle and vehicle-to-barrier simulations were generated such that the Acura experienced a change-in-velocity of 5 mph.

# 6 mph Delta-V Equivalents

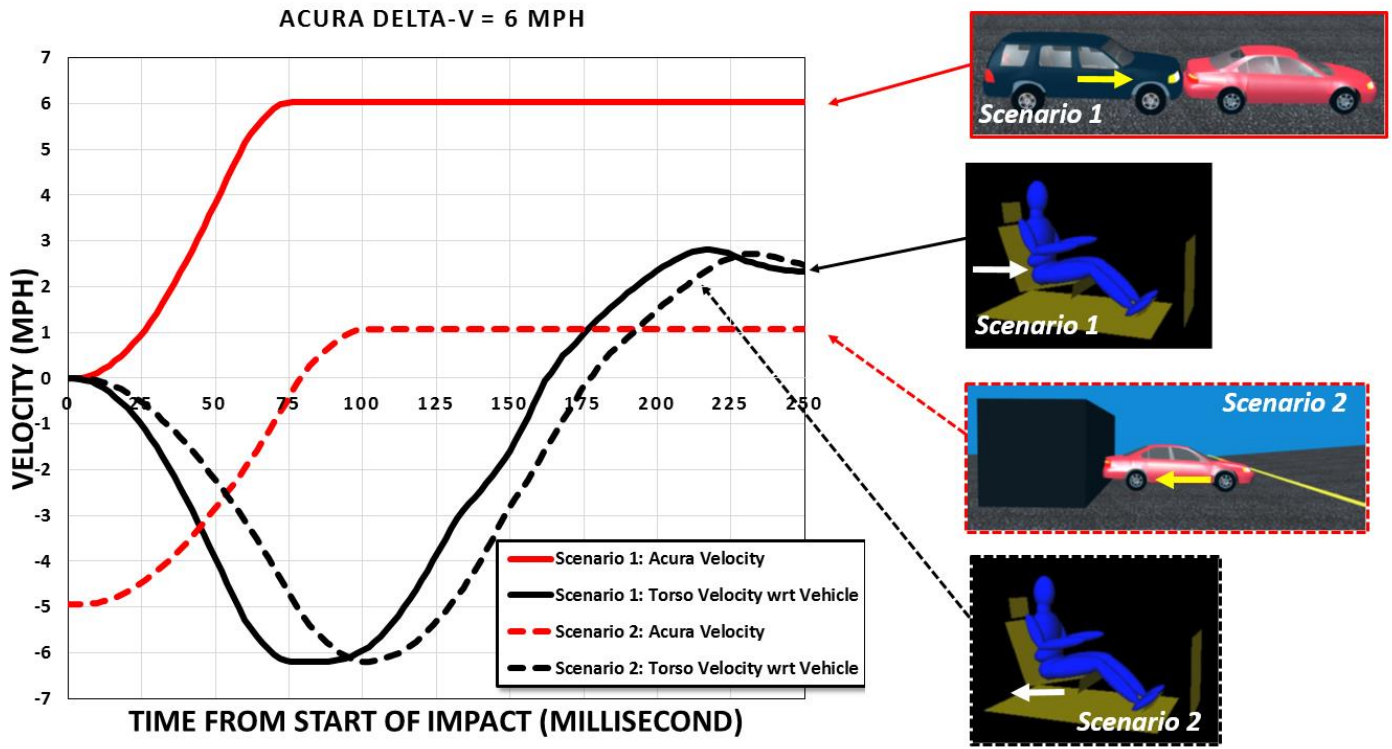

Figure 11: Vehicle and Occupant-Torso Velocities versus time from the start of impact. The vehicle-to-vehicle and vehicle-to-barrier simulations were generated such that the Acura experienced a change-in-velocity of 6 mph.

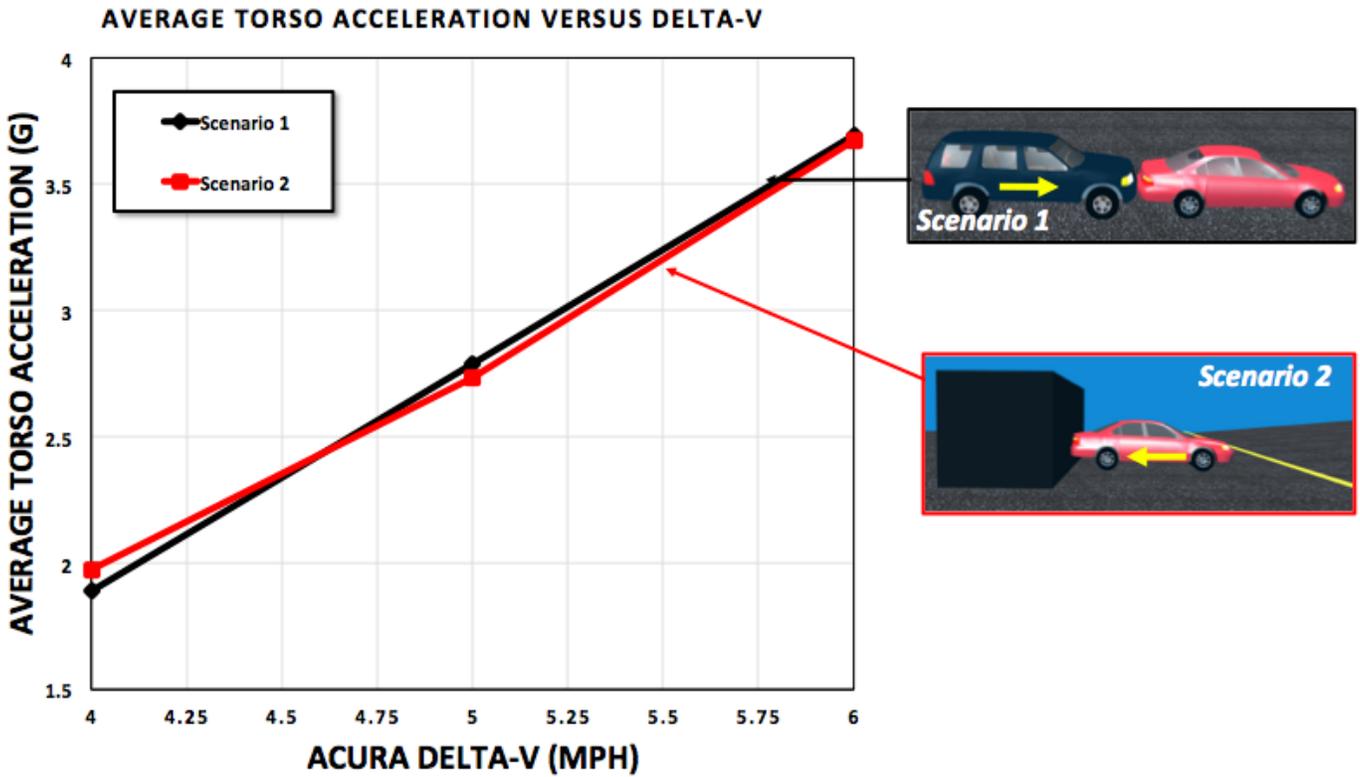

Figure 12: Average torso acceleration versus vehicle change-in-velocity for 4, 5, and 6 mph delta-V simulations. Black shows the results for vehicle-to-vehicle rear-impact collision simulations. Red shows the results from the equivalent barrier speed scenarios.

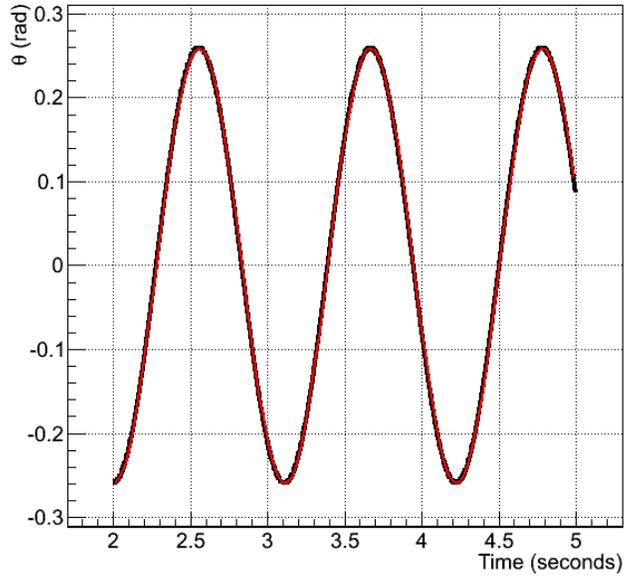

**Figure 13: Black: ROOT-based simulation output of ideal pendulum swing angle versus time. Red: Sinusoidal functional fit to simulation output.**

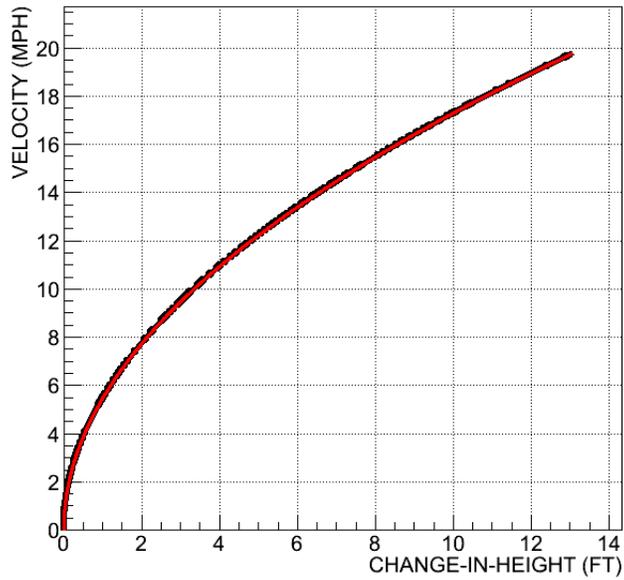

**Figure 14: Black: Output from a ROOT-based simulation of a vehicle undergoing pendulum motion starting from a 13 ft elevation in center-of-gravity height. Red: Work-energy based estimated velocity of the vehicle as it swings shown as a function of center-of-gravity change-in-height.**

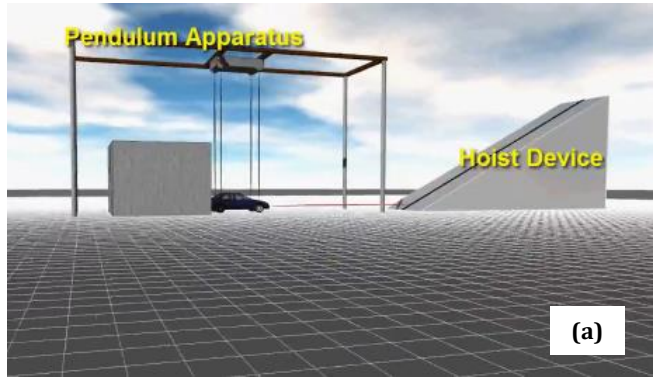

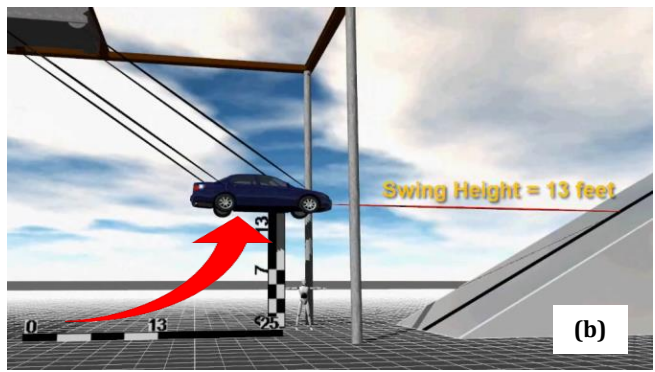

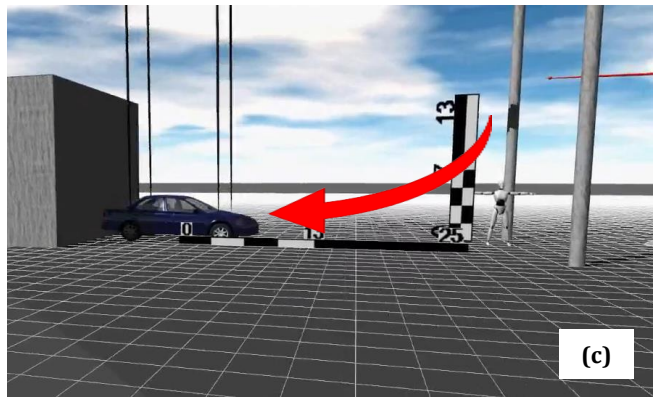

**Figure 15:** Screenshots from a pendulum impact animation. (a) Depicts the ideal structure of the bifilar pendulum apparatus, which consists of a simple hoist and a vehicle mounted and secured by cables to swing. The vehicle is made to impact a solid rigid barrier as it reaches the bottom of its arc. (b) Depicts the test vehicle being pulled up such that its center-of-gravity is displaced vertically by a distance of 13 ft. (c) Depicts the vehicle as it impacts a rigid barrier below with a velocity of 19.2 mph.